\documentclass{appolb}
\usepackage{epsfig}
\newcommand{\ave}[1]{\left\langle #1 \right\rangle}
\newcommand{\order}[1]{ \mathcal{O} \left( #1 \right) }

\newcommand{\dde}{\mathrm{d}\hspace{0.1em}}
 
  \newcommand{\lqcd}{\Lambda_{QCD}}

\begin{document}
\title{The phase diagram in $T-\mu-N_c$ space%
\thanks{Presented at ``three days on Quarkyonic island'', Wroclow, Poland}%
}
\author{Giorgio Torrieri$^a$, Stefano Lottini$^b$, Igor Mishustin$^a$, Piero Nicolini$^{a,b}$
\address{$^a$FIAS and $^b$ ITP, JW Goethe Universitat, Frankfurt, Germany}
}
\maketitle
\begin{abstract}
We examine the phase diagram of hadronic matter when the number of colors, as well as temperature and density, are varied.  We show that in this regime several phase transitions are possible, and we examine issues related to these transitions.
\end{abstract}
\PACS{PACS numbers come here}
  
\section{Introduction}

The large-$N_c$ limit \cite{thooft,witbar}, while sharing several qualitative characteristics with QCD, exhibits significant simplifications.   For this reason, it is a very popular technique for qualitative and semi-quantitative estimates of crucial QCD observables.   For instance, the string-gauge correspondence is profoundly intertwined with the large $N_c$ limit, since one of the requirements for a calculable dual holographic description is large $N_c$ \cite{adscft}.

Yet, the existence of a smooth limit between $N_c=3$ and $N_c = \infty$ is a conjecture which, for some observables, is {\em known} to be false.   We know that, at zero chemical potential, $N_c=\infty$ and finite $N_f$ deconfinement has to be a phase transition.  At $N_c =3,N_f=1,2$ lattice QCD seems to indicate it is a cross-over \cite{pollat}.   At finite chemical potential, in the confining phase, nuclear matter is known to be a strongly bound crystal in the large $N_c$ limit \cite{klebanov}.   It seems to be a liquid at $N_c=3$.  This means that, before predictions made at large $N_c$ can be made qualitatively reliable, the phase diagram in $T-\mu-N_c$ space needs to be explored.    The most current phenomenological application for these ideas is the existence of the so-called ``quarkyonic matter'' \cite{quarkyonic}.
This talk explores these ideas, summarizing \cite{mishustin,lottini,nicolini} and drawing connections between these works.
\section{The Van der Waals gas and large $N_c$ \label{secmishustin}}
Our first investigation of these issues \cite{mishustin} is to use the Van Der Waals (VdW) ansatz to see under what conditions can we obtain both the familiar nuclear matter phase diagram at $N_c=3$ and the commonly accepted large $N_c$ phase diagram at $N_c \rightarrow \infty$ \cite{quarkyonic}.   While the Van der Waals gas is certainly a very rough description, it is part of a universality class, to which the nuclear liquid-gas phase transition seems to belong.

In the large $N_c$ limit, the only $N_c$-invariant scale of the theory is $\lqcd$, the scale at which the 't Hooft coupling constant becomes $\lambda \sim \order{1}$.   While a precise value of this scale depends on the scheme used to calculate it, its roughly $\lqcd \sim N_c^0 \simeq 200-300$ MeV \cite{pdg}.
It is therefore natural to expect that any physical quantity is $\sim f(N_c) \lqcd^d$, a dimensionless function of $N_c$ times a power of $\lqcd$ set by the dimensionality $d$ of the quantity.
Henceforward we shall adopt this assumption, and, for brevity, set $\lqcd$ to unity in the equations.    
In this notation, the Van Der Waals parameters $a$, $b$ and the curvature correction become dimensionless $\alpha$,$\beta$,$\gamma$ times the appropriate power of $\lqcd$ (3 for $\alpha$,2 for $\beta$,4 for $\gamma$), and the VdW equation becomes
\begin{equation}
\left( \rho^{-1} - \alpha \right) \left(  P + \beta \rho^2 - \gamma  \rho^3 \right) = T
\label{vdw}
\end{equation}
We start by noting that, naturally, $\alpha$ can only go to $\lqcd^{-3}$ at $N_c \rightarrow \infty$, and is significantly larger than $\lqcd^{-3}$ at $N_c=3$.   Therefore, 
\begin{equation}
\label{eqalpha}
\alpha \sim \order{ \frac{N_N}{N_c}} + 1
\end{equation}
where $N_N$ is a constant to be determined.
The coefficients $\beta,\gamma$ should, according to \cite{witbar,klebanov} go as $N_c$.   Recent work \cite{larrynucleon}, however, has cast doubt on this assumption and proposed they go as $\sim N_c^0$ or $\sim \ln N_c$.

The chemical potential can be obtained by the textbook thermodynamic relation
$\rho = \left(dP/d\mu  \right)_T$.   Inverting, and writing in terms of $\mu_q=\mu_B/N_c$ we have
\begin{equation}
\label{mubexact}
\mu_q = 1 +\frac{1}{N_c} \left[ \int^\rho_0 f(\rho',T) d\rho' + F(T)  \right]
\end{equation}
where the first term is the nucleon mass and 
\begin{equation}
\label{fdef}
f(\rho,T)= \left( \frac{ dP}{d \rho} \right)_{T} \frac{1}{\rho} = \frac{T}{ \rho  (1-\alpha \rho )^2}+2\beta
\end{equation}
$\rho$ and $P$ are the density at the phase transition, which could be liquid  $\rho_l$ or gas $\rho_g$ (if the calculation is performed correctly the same chemical potential should come out).
$\rho_{l,g}$ are in turn the solutions to the equation \ref{vdw} in the region where this equation has two solutions (of which one is thermodynamically unstable).
Obtaining all such solutions is trivial at the mathematical level through algebraically cumbersome.  The reader can get the detailed results in \cite{mishustin}.

\begin{figure}[h]
    \centering
        \includegraphics[width=0.4\textwidth]{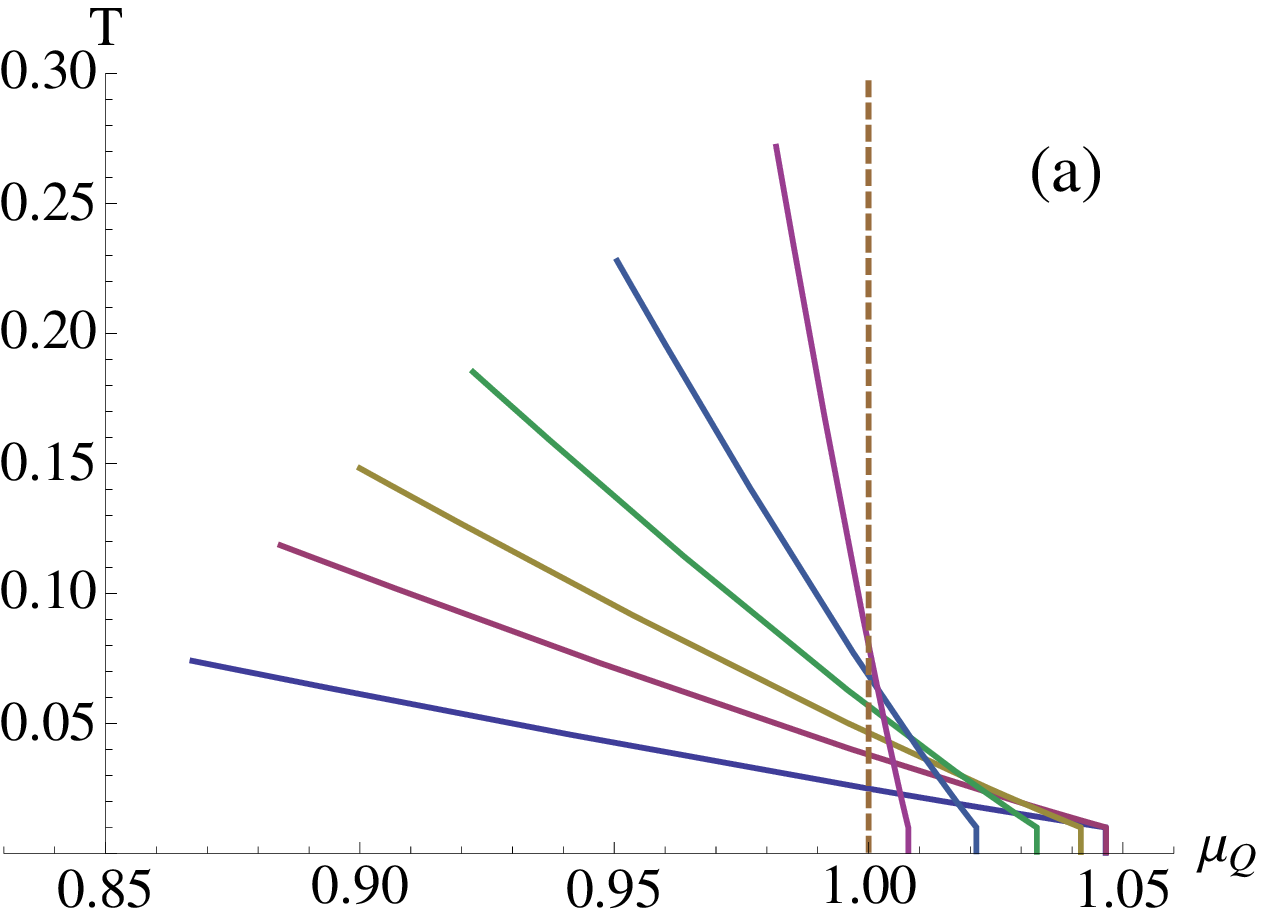}
\includegraphics[scale=0.6]{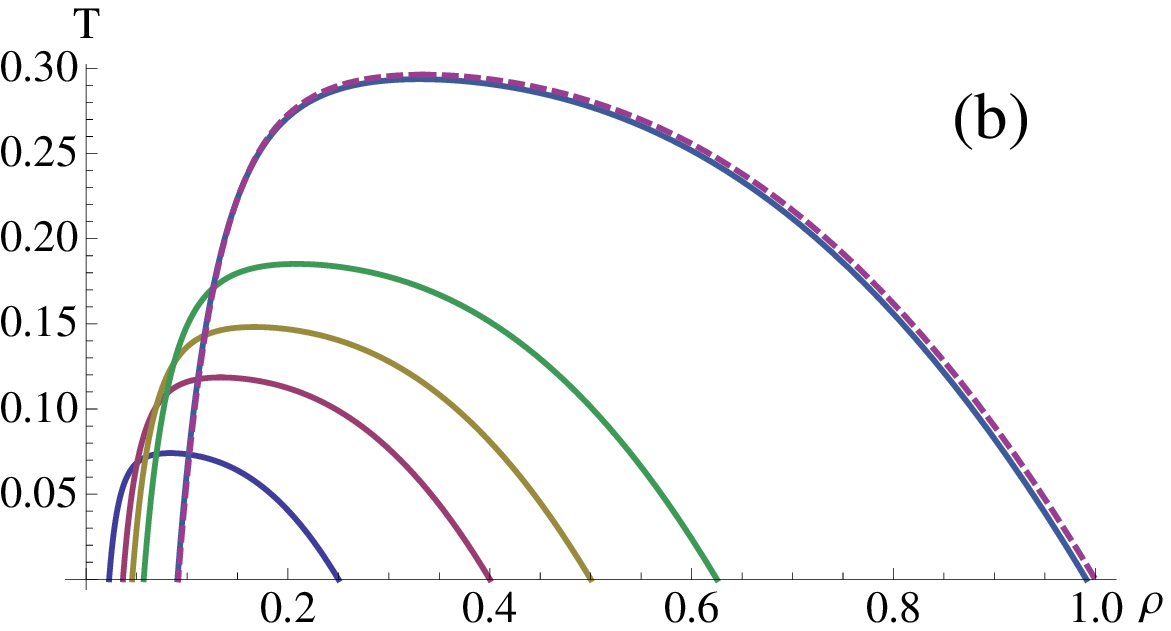}
       \includegraphics[width=0.4\textwidth]{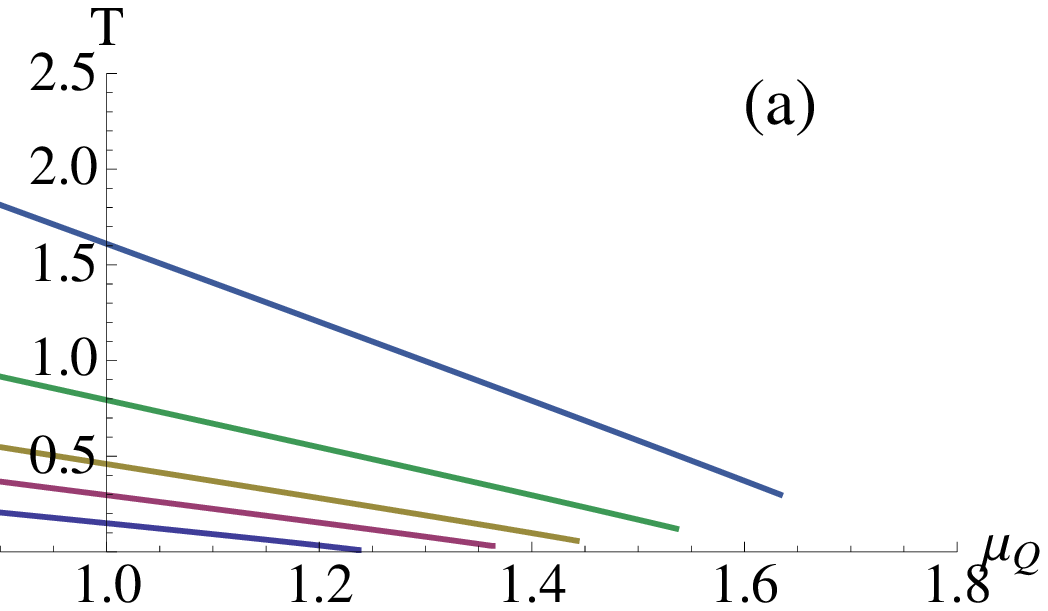} \includegraphics[scale=0.6]{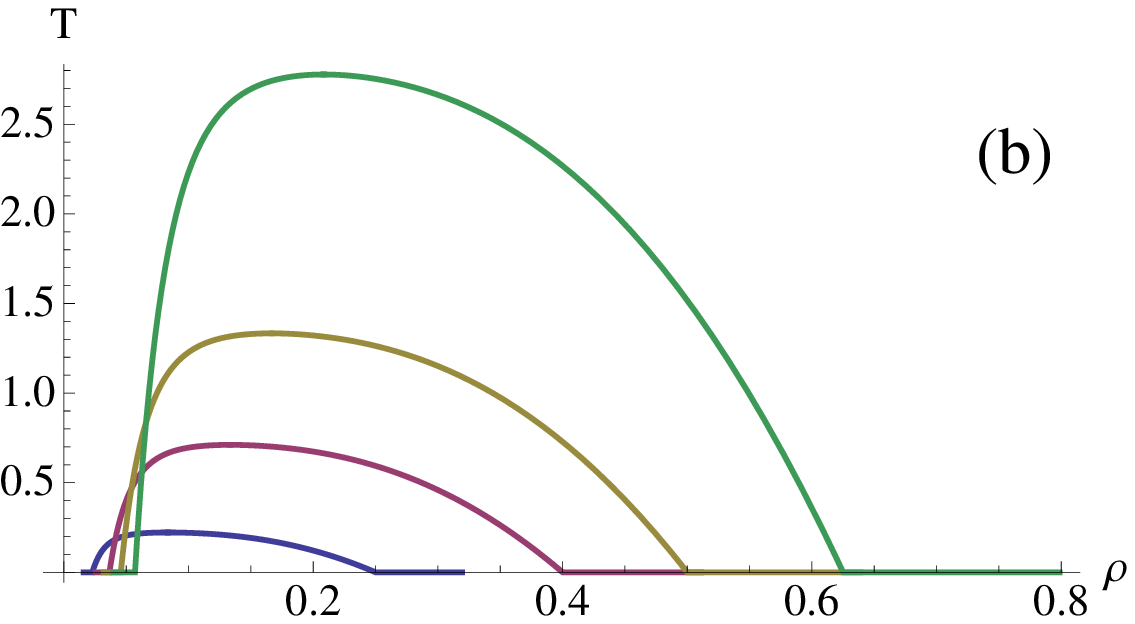} \caption{(color online)
The phase diagram in $T-\mu$ space (left panel) and $T-\rho$ space (right panel) as a function of $N_c$ ($N_c=3,5,8,10,30,100$, with increasing color
corresponding to a line with higher $T,\mu,\rho$) Top panels assume nuclear interactions $\sim N_c^0$ or $\sim \ln N_c$, bottom panels as $\sim N_c$
} \label{phasemu} \end{figure}

As can be seen (Fig. \ref{phasemu}), to interpolate between the currenlty accepted nuclear liquid phase diagram at $N_c=3$ ($\rho_c \ll \lqcd^3,T_c \ll \lqcd$) and the ``quarkyonic'' phase diagram at $N_c \rightarrow \infty$ ($\rho_c \sim \lqcd^3,T_c \sim \lqcd$), $N_N \sim \order{10}$ must be used.  How would we interpret physically such a value?   Above all, what is the physical meaning of $N_N$?
We conjecture that $N_N$  tracks another dimensionless scale relevant at high density:  The number of neighbors a nucleon has in a tightly packed nuclear material.  
 $N_N$,
of course, is a function not of $N_c$, but the (fixed) number of dimensions $d$ and ``packing scheme''.

The sensitivity of the varying $N_c$ equation of state to $N_N$ is undestandable due to the quantum nature of the problem and the fermionic nature of the quarks.
 The more neighbors, the more Pauli blocking of valence quarks must be important, and the more the presence of neighbors will disturb the configuration space part of the quark wavefunction inside the nucleons. 
Since,due to the uncertainty principle, any such disturbance of the nuclear wavefunction adds an energy of the order of the confinement scale $\sim \lqcd$, the nuclear repulsive core will be larger than the inverse of the nuclear separation up to the deconfinement temperature.  If the number of colors is larger than $N_N$, this problem will not exist since it will be possible to arrange the color part of the wavefunction so the nearest quarks of neighboring baryons will be of different colors. In this limit baryons can be tightly packed (interbaryonic separation $\sim \lqcd$) without the configuration space part of the baryonic wavefunction being disturbed.   

The number of neighbours scaling suggests that percolation phenomena might be relevant in the middle of the two regimes.   In the next section we investigate this possibility.

\section{Percolation and dense matter at varying $N_c$ \label{seclottini}}

The key insight suggesting that interesting structures might be lurking in $N_c$ is that 3D bond-percolation exhibits a phase transition at comparatively low critical link probability: for instance, $p_c\sim 0.25, 0.18, 0.12$ for simple-cubic, body-centered-cubic and hexagonal-close-packed lattice, respectively \cite{percthresholds}.
Such values suggest that long-distance correlations on the quark level could occur even with a somewhat low percentage of quarks hopping between baryons, i.e.~firmly in the confined phase.
While below $p_c$ the characteristic correlation distance $\xi$ ($\sim$ cluster size) is $\sim \lqcd^{-1}$, above the threshold this quantity explodes to the total system size in a comparable amount of time. We leave the meaning of ``correlation''  vague, as it can be either a quark hop or a gluon exchange; in our context, it implies exchange of colour degrees of freedom within a confined tightly packed medium.
We encode the likelihood of exchange between neighbouring baryons in a link probability $p$, to be compared with the percolation threshold $p_c$ in order to assess the formation of large-scale structures.

Two baryons will be correlated if {\em at least two} quarks are correlated.
One has to sum over all possible multi-quark configurations, resulting in a strong $N_c$-dependence of $p$.
We determine the latter by calculating the probability $q=1-p$ that {\em no} exchanges happen between neighbouring baryons.
Assuming the quarks inside the nucleon are uncorrelated (Fermi motion dominates), this probability factorises into a geometric distribution $f_{A,B}(\mathbf{x})$ for quarks to be at a certain (vector) position $\mathbf{x}$, and a ``squared propagator'' transition amplitude $F(d)$ for them:
\begin{eqnarray}
		\label{ptotlink}
	p &=& 1- \big(q_{(1),ij}\big)^{(N_c)^\zeta}\,;\\
	q_{(1),ij} &=& 
		\int  f_A (\mathbf{x}_i) \dde \mathbf{x}_i \int f_B(\mathbf{x}_j) \dde \mathbf{x}_j
		\left( 1- F(|\mathbf{x}_i-\mathbf{x}_j|)  \right)  \;. \nonumber
\end{eqnarray}
We assume a ``hard-sphere'' distribution for $f_{A,B}$
(since we keep $\mu_q$ fixed, the distance between centres of neighbouring baryons is always $2\lqcd^{-1}$)
\begin{equation}
f_{A,B} (\mathbf{x}) \;\; \propto \;\;  \Theta \left( 1  - \lqcd \left| \mathbf{x} - \mathbf{x}_{A,B}^{\mathrm{centre}} \right| \right)
\end{equation}
 and a probability of exchange $i\leftrightarrow j$ based on a range of ``reasonable'' propagators, compatible with confinement (fast fall-out in configuration space at distances greater than $r_T \sim 1$ in units of $\lqcd^{-1}$) and with the large $N_c$ limit of QCD, the interaction is 
$\sim g^2 \sim \lambda/N_c$ \cite{thooft,witbar}.
The propagators we use are the simple $\Theta$-function in configuration space and the momentum-space $\Theta$-function used in \cite{kojo}, all normalised 
so their area is $\lambda r_T/N_c$ where $r_T$ is the range of propagation ($\sim \lqcd^{-1}$).  In configuration space the transition amplitudes are, respectively,
\begin{equation}
\label{propagator}
F(y) = \frac{\lambda}{N_c}  \left\{ 
\begin{array}{c}
\Theta(1 - \frac{y\lqcd}{r_T})\\
 \frac{2 r_T^2 }{\pi y^2} \sin^2\Big(\frac{y\lqcd}{r_T}\Big)
\end{array}
 \right.
\end{equation}
Other transition amplitudes, such as a Gaussian distribution in configuration space, were also tried with no significant modifications.  This is unsurprising since, due to the fact that 3D percolation has a second-order phase transition at a certain $p_c < 1$, the results we obtain below have some 
degree of universality: as long as the qualitative features of confinement are observed (the transition amplitude $F(y)$ drops sharply above the scale $r_T$, and the hadron density profile $f_{A,B}(\mathbf{x})$ has a central plateau of radius $\sim \lqcd^{-1}$ and a sharply decreasing
tail outside), the results we show vary quantitatively but not qualitatively.

The crucial parameter left is $\zeta$ in Eq.~\ref{ptotlink}.
One can easily see that $\zeta = 1$ is in contradiction with the Skyrme crystal picture at large $N_c$:  in this picture, $p(N_c)$ approaches a constant large-$N_c$ value {\em from above}: {\em low $N_c$} nuclear matter would be more correlated (and hence more strongly bound) than {\em high $N_c$} nuclear matter.   Comparing strongly coupled $N_c \rightarrow \infty$ nuclear matter \cite{witbar} to the weakly bound nuclear liquid at $N_c=3$ \cite{mishustin}, this is obviously not right.
We therefore assume $\zeta=2$ henceforward, natural if the link is actually realised by a gluon exchange rather than a quark flip (it is obvious this does indeed dominate at large $N_c$).

In the large $N_c$ limit for the case $\zeta=2$, $p$ asymptotically approaches unity.
It is reasonable that this is the point where the ``dense baryonic matter as a Skyrme crystal'', theorised in \cite{witbar,klebanov,quarkyonic}, is reached.
If this is the case, however, one should remember that a percolation second-order phase transition occurs at a $p_c\ll 1$. 
Hence, keeping $\mu_q \sim \lqcd$ fixed but varying $N_c$, the features of the Skyrme crystal should manifest not with a continuous approach, rather as a second order transition at a not too high $N_c$, whose order parameter can be though to be the ``giant cluster'' density.
Below the critical $N_c$ there is little correlation between quarks of different baryons, while above this threshold they can correlate, with the distance boundary given only by causality.
We reiterate that this is {\em not} deconfinement since $\mu \sim \lqcd \ll N_c^{1/2} \lqcd$ independently of the number of colours, and the {\em fraction} of correlated quarks from different hadrons is still $\sim 0.1-0.3 \ll 1$ at the percolation transition. Right above this transition, therefore, the baryonic wavefunction should {\em not} be too different from the large-$N_c$ baryonic wavefunction described in \cite{witbar}. The {\em correlation distance} of quarks will however be much larger than the baryon size.
The features of this new phase are therefore similar to those of the quarkyonic matter \cite{quarkyonic}.

\begin{center}
\begin{figure}[t]
\epsfig{height=4cm,figure=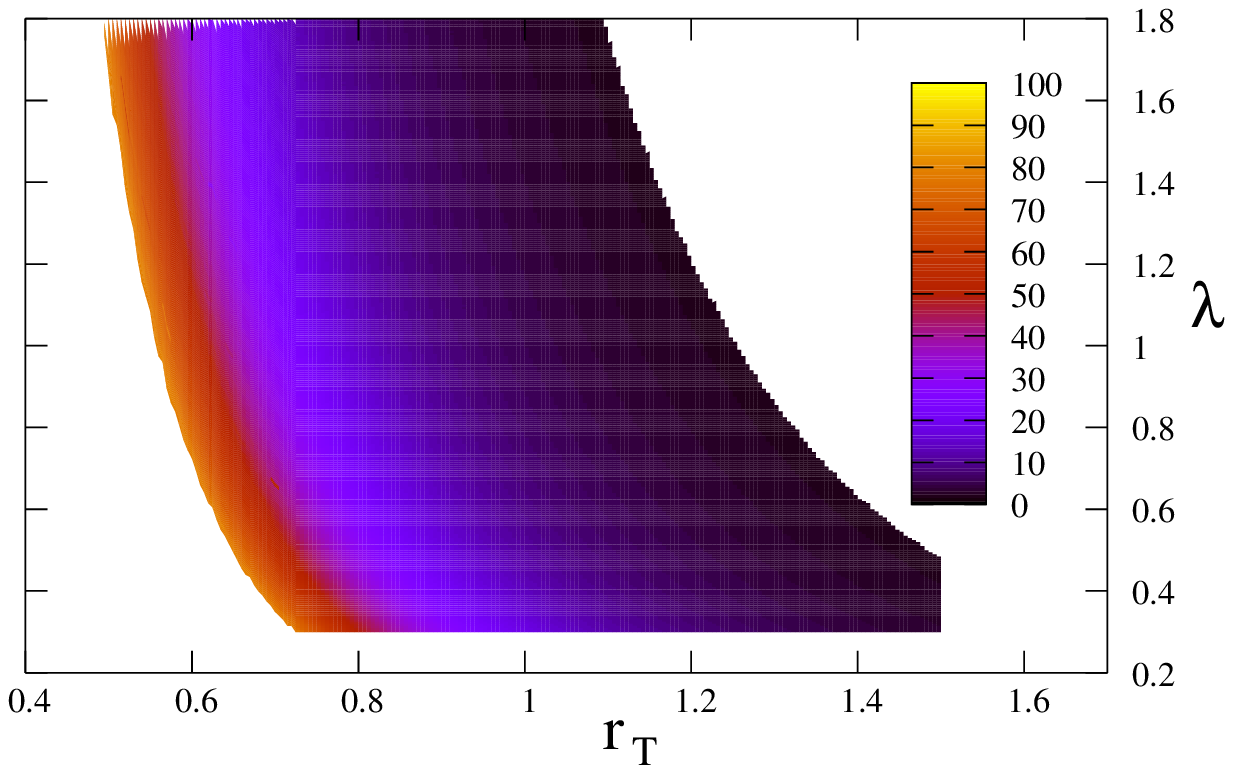}
\epsfig{height=4cm,figure=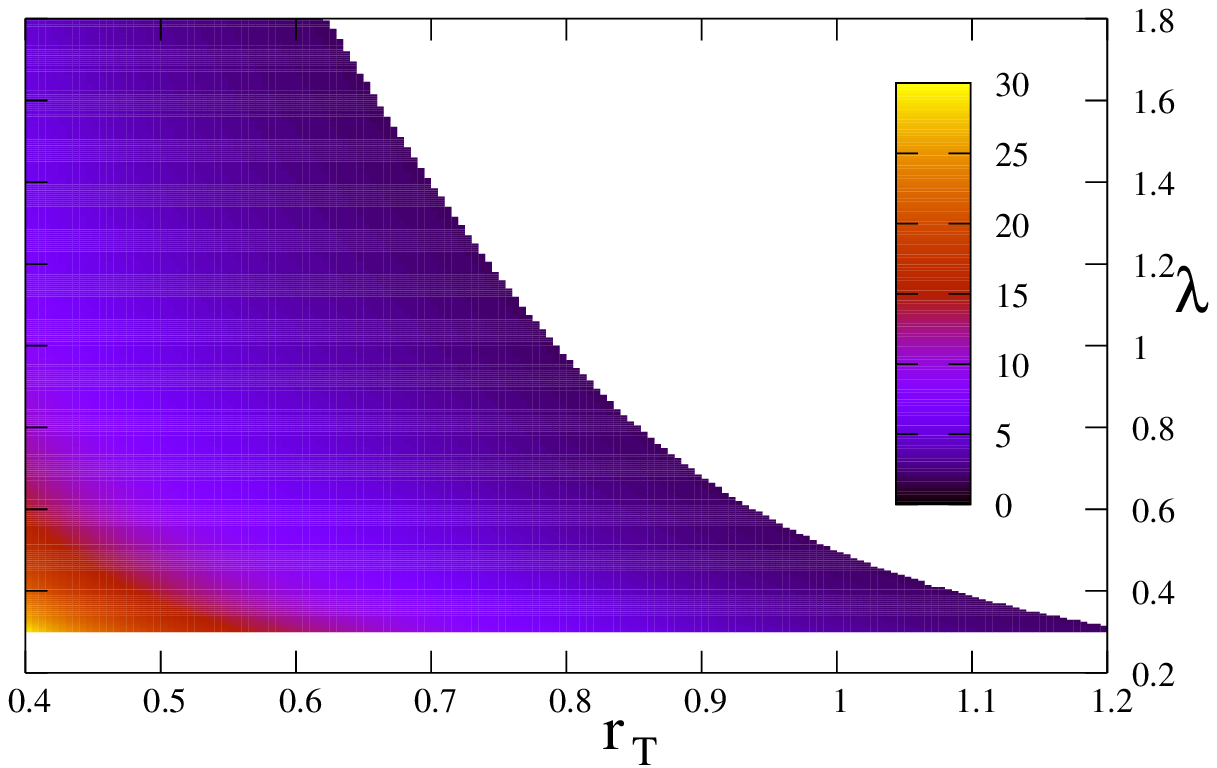}

\caption{\label{lottini} (color online) Contour plot of the critical $N_c$ for the percolation transition in a hexagonally packed lattice as a
function of the coupling $\lambda$ and range $r_T$ (in $\Lambda_{QCD}$ units). The left panel assumes a $\Theta$-function correlation probability in
configuration space, the right panel assumes a correlation probability based on the propagator used in \cite{kojo}.  Diagram covers $2 \le N_c \le
80$.  See \cite{lottini} for details
}

\end{figure}
\end{center}

Assuming the lowest 3D value $p_c=0.12$, appropriate for a closely packed hexagonal lattice, the critical number of colours is shown in Fig.~\ref{lottini} as a function of $\lambda$ and $r_T$.  As can be seen, the critical number of colours is significantly larger than three for $r_T \sim \lqcd^{-1},\lambda \sim 1$.  
Considering Fig.~\ref{lottini} is a {\em lower limit} since $p_c$ is at its minimum ($p_c$ is significantly higher both in a Skyrme cubic crystal and in a disordered fluid), we can  say that $N_c=3$ is disfavoured, although it can not be excluded.
Changing temperature and $\mu_q$ should further change the critical $N_c$.   Exploring this parameter space, and seeing how it relates to the confinement phase transition, is the subject of a forthcoming work.

What are the phenomenological consequences of percolation?
If by ``correlation'' we mean energy-momentum-exchange via quark tunneling between baryons, it is reasonable that pressure and entropy density $\sim N_c$ above the percolation  threshold, while below it they stay $\sim N_c^0$.  This is because above the threshold, where interbaryon tunnelling is significant, "typical" excitations of the Fermi surface will be
 superpositions across baryons of baryon-localized quark-hole excitations, similar of conduction band electrons in a metal; while the localized excitation energy $\geq \lqcd$, the superposition makes its energy $\sim \lqcd$ even if color degeneracy remains. 
 Thus, the degrees of freedom of the system above percolation will be delocalized weakly interacting quarks in a lattice of confining potentials, a picture compatible with \cite{quarkyonic}.
Below the threshold, where tunnelling is negligible, excitations are either color singlets or of energy $E \gg \lqcd$, suppressed below deconfinement.

\section{Confinement at finite $N_c$ and quantum gravity \label{secnicolini}}
The transitions discussed here might also be visible with holographic techniques {\em beyond} the supergravity limit, since finite $\lambda N_c^{-1}$ corresponds, 
in gauge/string duality, to the string coupling constant $g_s$ \cite{adscft}. 
Thus, a transition at finite $N_c$ will manifest itself as a transition between a classical supergravity regime and a quantum-gravitational regime.
Since gravity to one loop still has not been solved for backgrounds dual to confining theories, we can not rigorously say anything further.

However, in \cite{nicolini}, we have made a first step in that direction using the reasonable  Ansatz, developed earlier \cite{nicolreference,nicolreview}, that the effect of Quantum gravity corrections is to smoothen the Black hole singularity to a Gaussian of with $\ell$, the Plank length.   The problem of the first order quantum black hole reduces, than, to the problem of the classical general relativity sourced by a Gaussian distribution of energy density $\rho$.  The top energy-momentum tensor is then
 \begin{equation}
 T_0^0=-\rho_\ell (r)=-\frac{M}{\left(4\pi\ell^2\right)^{3/2}}\,\exp\left(-\frac{r^2}{4\ell^2}\right).
 \label{tgaussian}
\end{equation}
with the other components of $T^{\mu \nu}$ being fixed by isotropy, the classical Einstein equation
\begin{equation}
 R_{\mu nu} - \frac{1}{2} R g_{\mu \nu} - \frac{3}{L_{ADS}^2} g_{\mu \nu} = 8 \pi G T_{\mu \nu}
\end{equation}
 and conservation laws ( $T^{\mu,\nu}_{;\mu}=0$).  Thermodynamics can be obtained by the usual relationship between the entropy $s$ and the horizon area $A$, $s=A/G$, as well as Maxwell's relations \cite{hawkingpage}.

We rely on \cite{nicolini,nicolreference} for details on solving this problem, and connecting it to the Hawking-Page phase transition \cite{hawkingpage}, and just give the result: 
\begin{equation}
\label{freen}
F(r_H) = F_{hawking}(r_H ) + \Delta F(r_H )
\end{equation}
where $F_{hawking}$ is the free-energy of a classical black hole in AdS space \cite{hawkingpage} and \[\ \Delta F(r) = 2 \pi T  \int_{r_0}^{r_H} \frac{r dr}{G\frac{2}{\sqrt{\pi}}\gamma(3/2; r^2/4\ell^2)} \]

\begin{figure}
 \begin{center}
  \includegraphics[height=5.5cm]{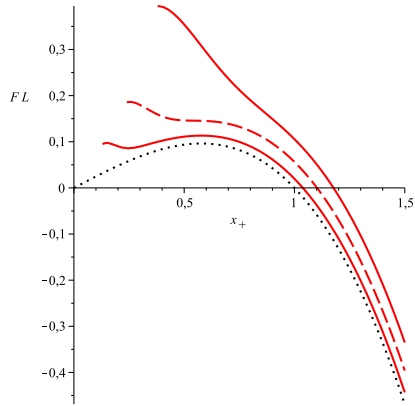}
      \caption{\label{nicolini}The free energy $F$ in units of the cosmological constant $L$ for the system in equilibrium as a function of the 
Schwarzschild radius $x_+$ from top to bottom for, in terms of $q=\ell/L$, $q=1/3$ and $q=0.1$ (solid lines). The dashed lines represent the critical
case $q=q^\ast$, while the dotted line is the classical Hawking-Page phase transition ($q=0$). See \cite{nicolini} for details}
     \end{center}
  \end{figure}

Once we examine the minimum structure of this free energy, we find a critical point:
For a ``critical''  $q = \ell/L_{ADS} \simeq 0.18$  there is a critical point after which the Hawking-Page phase transition becomes a {\em cross-over} (Fig. \ref{nicolini}).   The resulting phase diagram (in $T$ vs $\ell$) is shown to be in the universality class of the Van der Waals gas.

The relevance of this paper to the topic at hand becomes apparent given that the Hawking-Page transition is thought to be dual to the deconfinement transition \cite{adscft}.   We know that at finite $N_f/N_c$, deconfinement becomes a cross-over rather than a transition.   Thus, this calculation can be thought of as an estimate of the critical $N_f/N_c$ for the deconfinement critical point in two-dimensions (since the gravitatational setup is in 3D).

While the $AdS_3$ case presented here is considerably different from the $AdS_1 \times S_n$ or the $AdS_5 \times S_5$ background examined in \cite{adscft}, our results fall neatly into the Van Der Waals universality class.
Because of that, and the structure of the calculations in the previous section, we expect that the general structure of this phase diagram will be maintained in $AdS_1 \times S_n$ or any other background where a deconfinement phase transition exists.
The critical $q^*$ will of course be different, but not the behavior around the critical point.       This critical point will translate itself in the behaviour of $N_c$ of the boundary theory, once again in a way that is universal around the critical point but sensitive to the exact nature of the background theory.

When the number of colors $N_c \gg$ the number of flavors $N_f$ (the so-called T'Hooft limit), confinement is rigorously known to be a phase transition in all dimensions greater than one, since the symmetries of the system are expected to change. The low temperature phase will have acquired an extra $Z_N$ symmetry, signifying a zero expectation value of the Polyakov loop \cite{poldef}.
In the high-temperature phase the Polyakov loop will aquire a finite expectation value, and $Z_N$ will be spontaneusly broken.    In the gravity picture, this can be thought of as a manifestation of the widely-believed cosmic censorship conjecture 
\cite{penrose}, the idea that any singularity in spacetime is surrounded by an event horizon (and hence a discontinuity in the energy density at the boundary theory).   Since the existance of the black hole tracks the confining Polyakov loop behaviour in \cite{adscft}, making confinement into a cross-over would mean smoothening the black hole singularity.    This can only be done, eg, by our ansatz, for $q >0$.

At finite number of flavors $N_f$, the $Z_N$ symmetry is no longer exact at a fundamental level.  Hence, it is natural to expect that at some critical $N_c^{crit}$ ``a critical point'' for deconfinement appears, where for higher $N_c$ deconfinement is a phase transition and for lower $N_c$ it is a crossover.  This is indeed what emerges from lattice simulations \cite{pollat} in three dimensions with a finite number of flavors.   The free energy with respect to the Polyakov loop expectation value $\ave{L}$ and energy $E$ is therefore of the form
\begin{equation}
\label{freengauge}
F\left( \ave{L},E \right) = F_{gauge} \left( \ave{L},E \right) +  \delta F \left(\frac{N_f}{N_c}, \ave{L},E \right).
\end{equation}
$F_{gauge}$ at the mixed phase is a function with two local minima, corresponding to the two phases.  For high-enough $N_f/N_c$, however, $\delta F$ could destroy one of the minima and make the phase transition into a crossover.
The Van der Waals form of the free energy Eq. \ref{freen} exactly parallels Eq. \ref{freengauge}, with the $q$ parameter playing the role of $N_f/N_c$ in a Gauge theory in 2+1 dimensions (since the bulk calculated here is 3+1 dimensions). 

\section{Discussion and Conclusions}
The preceding three sections, at first sight, look like they deal with very different subjects.  Indeed, the current status is that any correlation between them is conjectural.   

When $N_c$ is varied in the dense matter strongly coupled regime, deviations from the ``classical'' skyrme crystal arise due to the interplay between the number of colors and the number of neighbours $N_N$ in the densely packed system.  This ``geometrical'' picture is explored rigorously, in terms of an analytical model, in section \ref{seclottini} and shown to give a percolation-type phase transition.   A similar behaviour can be argued in terms of the Pauli exclusion principle, and if {\em it is assumed}, as shown in section \ref{secmishustin}, leads to the transition between the loosely-bound nuclear liquid at $N_c=3$ and the tightly bound nucelar crystal at $N_c \gg N_N$.
Since the solid-liquid transition can be understood as a percolation of crystal bonds \cite{solid}, it is therefore compelling to identify the percolation transition with the physics described in section \ref{secmishustin}, with the critical $N_c$ playing the role of ``critical point'' in $T-\mu-N_c$ space (the transition in section \ref{seclottini} is necessarily second order).

The connection with the Hawking-Page transition described in section \ref{secnicolini} is, at the moment, not so clear.   It is worth pointing out, however, that this transition is also  in the Van der Waals universality class.   While section \ref{secnicolini} was at zero chemical potential (the black hole was not charged), the Hawking-page phase diagram for classical ($\ell=0$) charged black holes also falls into this class \cite{emparan}.   

Thus, the phenomena examined at finite chemical potential in section \ref{seclottini} and \ref{secmishustin} might be analogous to the dynamics examined in section \ref{secnicolini} at zero chemical potential.  The closeness of the critical $q^*$ to the  critical number of $ N_f/N_c$ required for percolation in 2D \cite{lottini} (6 colors per flavor) is in this sense intriguing.   

Physically, in $AdS_3$, the Hawking-Page transition coincides with the transition from a gas of black holes to one large black hole. It is plausible that the critical point to a Hawking-Page cross-over occurs because the black holes in the gas are actually linked by super-horizon quantum gravitational interactions, modeled in our approach by
``smearing'' the super-horizon black hole distance by a Planck-sized width.  In this scenario, the connection of percolation to the cross-over in deconfinement is straight-forward.   Before this connection can be made more rigorous, however, a Gauge/string setup where percolation is relevant must be examined with a ``smearing'' analogous to section \ref{secnicolini}.
Work of this kind is in progress.   

IM acknowledges support provided by the DFG grant 436RUS 113/711/0-2 (Germany) and grant NS-7235.2010.2 (Russia). G.T. acknowledges the financial
support received from the Helmholtz International Center for FAIR within the framework of the LOEWE program (Landesoffensive zur Entwicklung
Wissenschaftlich-\"Okonomischer Exzellenz) launched by the State of Hesse.
We acknowledge the organizers of the Max Born symposium for their generous support and for the very stimulating atmosphere at this workshop.

\end{document}